\documentclass[reprint, superscriptaddress,
 amsmath,amssymb,
 aps, 
floatfix,]{revtex4-2}

\usepackage{graphicx}
\usepackage{dcolumn}
\usepackage{bm}
\usepackage{lipsum}
\usepackage{xcolor}

\usepackage{float}

\makeatletter
\def\@fnsymbol#1{\ensuremath{\ifcase#1\or \ddagger\or *\or \dagger\or
   \mathsection\or \mathparagraph\or \|\or **\or \dagger\dagger
   \or \ddagger\ddagger \else\@ctrerr\fi}}
\makeatother
    
\begin{document}
\preprint{APS/123-QED}
\title{Controlling nonlinear interaction in a many-mode laser by tuning disorder}
\author{Yaniv Eliezer}
\thanks{These authors contributed equally to this work.}
\affiliation{Department of Applied Physics, Yale University, New Haven, Connecticut 06520, USA}
\author{Simon Mahler}
\thanks{These authors contributed equally to this work.}
\affiliation{Department of Physics of Complex Systems, Weizmann Institute of Science, Rehovot 761001, Israel}
\author{Asher A. Friesem}
\affiliation{Department of Physics of Complex Systems, Weizmann Institute of Science, Rehovot 761001, Israel}
\author{Hui Cao}
\affiliation{Department of Applied Physics, Yale University, New Haven, Connecticut 06520, USA}
\author{Nir Davidson}
\email{nir.davidson@weizmann.ac.il}
\affiliation{Department of Physics of Complex Systems, Weizmann Institute of Science, Rehovot 761001, Israel}
\date{\today}

\begin{abstract}
A many-mode laser with nonlinear modal interaction could serve as a model system to study many-body physics. However, precise and continuous tuning of the interaction strength over a wide range is challenging. Here, we present a unique method for controlling lasing mode structures by introducing random phase fluctuation to a nearly degenerate cavity. We show numerically and experimentally that as the characteristic scale of phase fluctuation decreases by two orders of magnitude, the transverse modes become fragmented and the reduction of their spatial overlap suppresses modal competition for gain, allowing more modes to lase. The tunability, flexibility and robustness of our system provides a powerful platform for investigating many-body phenomena. 
\end{abstract}
\maketitle

Many-body interaction has been a general topic in numerous fields of research, including condensed matter and particle physics, astronomy, chemistry, biology, neuroscience, and even social sciences. In optics, many-body interactions have been studied in a variety of active systems with different types of nonlinearities, which display a range of phenomena such as synchronization, pattern formation, bistability and chaotic dynamics~\cite{ARECCHI19991}. A well-known example is multimode lasers, where many lasing modes interact nonlinearly through the gain material. The complex interactions provide an optical realization of XY spin Hamiltonian and geometrical frustration~\cite{PhysRevResearch.2.033008, nixon2013observing, parto2020realizing}. In a random laser, the nonlinear coupling of lasing modes in a disordered potential leads to the ``glassy'' behavior and a replica-symmetry breaking phase transition~\cite{angelani2006glassy, antenucci2015general, ghofraniha2015experimental}.  

A continuous tuning of modal interaction strength over a wide range is essential to investigate many-body interaction, but it is difficult to realize experimentally. Previously, spatial modulation of pump intensity (optical gain) was adapted for controlling nonlinear interaction of lasing modes in random media~\cite{PhysRevLett109033903-2012, leonetti2012tunable, leonetti2013switching, leonetti2013active, bachelard2014adaptive}. While the lasing modes compete for optical gain, the degree of competition depends on the spatial and spectral overlap of these modes~\cite{cao2003mode, andreasen2014partially, ge2014enhancement}. Tuning the amount of disorder can vary the spatial distribution of random lasing modes, modifying their overlap~\cite{vanneste2001selective, vanneste2007lasing}. However, the lasing thresholds of these modes are also changed, in correspondence to the changes in their lifetimes or quality ($Q$) factors~\cite{wu2004random}. As the number of lasing modes varies, their interaction through gain saturation is affected. Therefore, it would be desirable to tune the spatial overlap of the lasing modes without significant modification of their thresholds.   

In this Letter, we introduce transverse disorder to a self-imaging cavity thereby inducing fragmentation of lasing modes. By varying the spatial scale of random phase modulation imposed by a spatial light modulator inside a degenerate cavity, we gradually tune the transverse size of lasing modes over two orders of magnitude. As the lasing modes adapt to the random phase variations and become localized in separate domains, their spatial overlap is reduced, and their nonlinear interaction via gain competition is suppressed. Unlike with random lasers, the $Q$ factors of many modes are determined mainly by the longitudinal confinement which remains constant during the tuning of transverse disorder, allowing these modes to lase simultaneously. Experimentally, the number of lasing modes increases as the characteristic length scale of random phase fluctuation decreases, indicating that the reduction of nonlinear modal interaction dominates over $Q$ factors spoiling.

The increase in the number of lasing modes due to modal fragmentation by disorder bears a resemblance to the fragmentation of Bose-Einstein condensates (BEC) with repulsive interactions in a disordered potential~\cite{modugno2010anderson}. The energy cost of fragmentation, proportional to spatial overlap of fragmented BECs~\cite{RevModPhys73307_2001}, is suppressed as the BECs become localized by the disordered potential, similarly to the cost of gain competition suppressed for the localized lasing modes. The mapping between energy cost in atomic systems and gain/loss in photonics~\cite{nixon2013observing} can therefore be used to study other many-body interacting systems, in particular the interplay between nonlinear interaction and disorder, using photonic simulators~\cite{mcmahon2016fully, tradonsky2019rapid}.

\begin{figure}[hbtp]
\centering
\includegraphics[clip, trim = 0.0cm 0.0cm 0cm 0cm, width=1\linewidth]{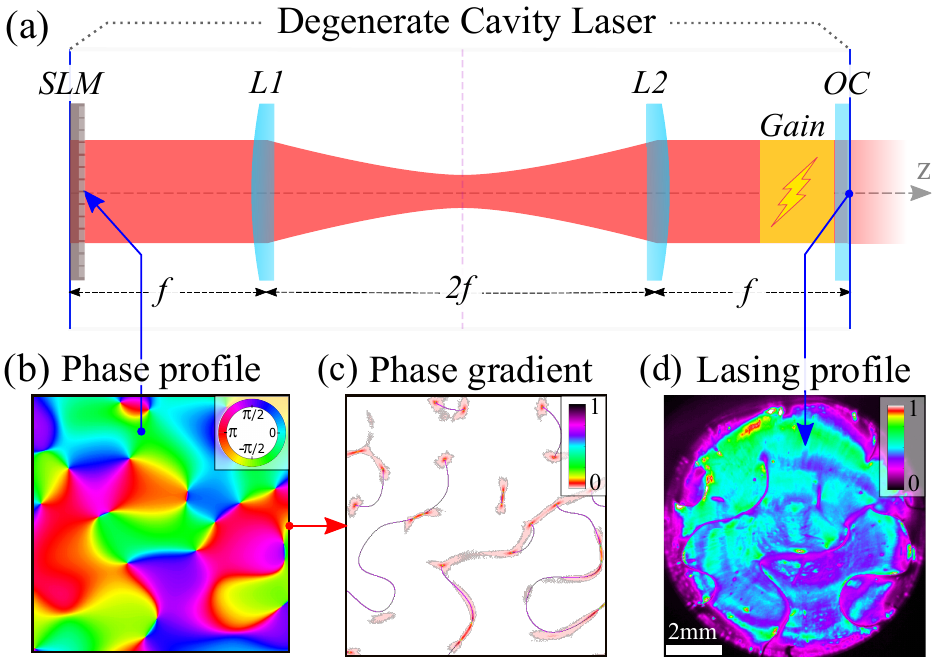}
\caption{{Introducing disorder to a degenerate cavity laser}. {(a)} Schematic of a degenerate cavity laser (DCL), comprised of a spatial light modulator (SLM), two lenses (L1, L2), and an output coupler (OC). The $4f$ configuration ensures a self-imaging condition. {(b)} A computer-generated random phase profile $\phi_{\xi}(x,y)$, with correlation length $\xi$~=~1.5~mm, is written to the phase-only SLM. {(c)} Calculated phase gradient of the profile in (b). {(d)} Experimentally measured emission intensity distribution on the OC at the pumping level of 2.2 times the lasing threshold for flat phase. The intensity nearly vanishes along the high-phase-gradient contours shown in (c), which effectively segment the emission pattern.}
\label{fig:setup}
\end{figure}

Figure~\ref{fig:setup}(a) is a schematic of our degenerate cavity laser (DCL). It is comprised of a reflective spatial light modulator (SLM), a Nd:YAG rod optically pumped to provide gain, a pair of lenses (L1, L2) arranged in a $4f$ configuration, and an output coupler (OC). The telescope formed by L1 and L2 images the SLM surface onto the OC and then back to the SLM~\cite{Supplemental}. The self-imaging condition allows many transverse field distributions to be eigenmodes of the cavity. The typical DCL has a flat mirror in place of the SLM, and it has many transverse modes with nearly degenerate frequency and loss~\cite{Arnaud69}. By inserting a SLM into the degenerate cavity, the transverse mode structure may be reconfigured easily and arbitrarily~\cite{Chene2021DDCL}. A computer-generated random phase profile $\phi(x,y)$ is displayed on the SLM. The random phase spatial correlation function 
\begin{equation}
    C_\phi(\Delta x, \Delta y) = \langle \phi(x, y) \phi(x+ \Delta x, y+ \Delta y) \rangle_{x, y} 
\end{equation}
is computed, where $\langle ... \rangle_{x, y}$ denotes averaging over the spatial coordinates $x$ and $y$. Its full width at half maximum (FWHM) gives the correlation length $\xi$ of phase fluctuation~\cite{Supplemental}. The SLM enables continuous tuning of $\xi$ from 0.1 mm to 10 mm, providing more flexibility over a glass phase diffuser with a constant $\xi$~\cite{Mahler2021}. 

Figure~\ref{fig:setup}(b) shows an example of a random phase profile displayed on the SLM with $\xi$~=~1.5~mm, and Fig.~\ref{fig:setup}(c) shows the corresponding phase gradient. The contours of large phase gradients reflect rapid phase variations, which lead to strong optical diffraction. Figure~\ref{fig:setup}(d) is the measured lasing emission pattern at the OC plane, which corresponds to the random phase profile at the SLM. As evident, the emission intensity drops abruptly along the high-phase-gradient contours, indicating that the lasing modes avoid these regions with high diffraction loss. Consequently, the lasing modes are segregated by the random phase profile.

\begin{figure}[hbtp]
	\centering
	\includegraphics[clip, trim = 0 0cm 0cm 0,width=1\linewidth]{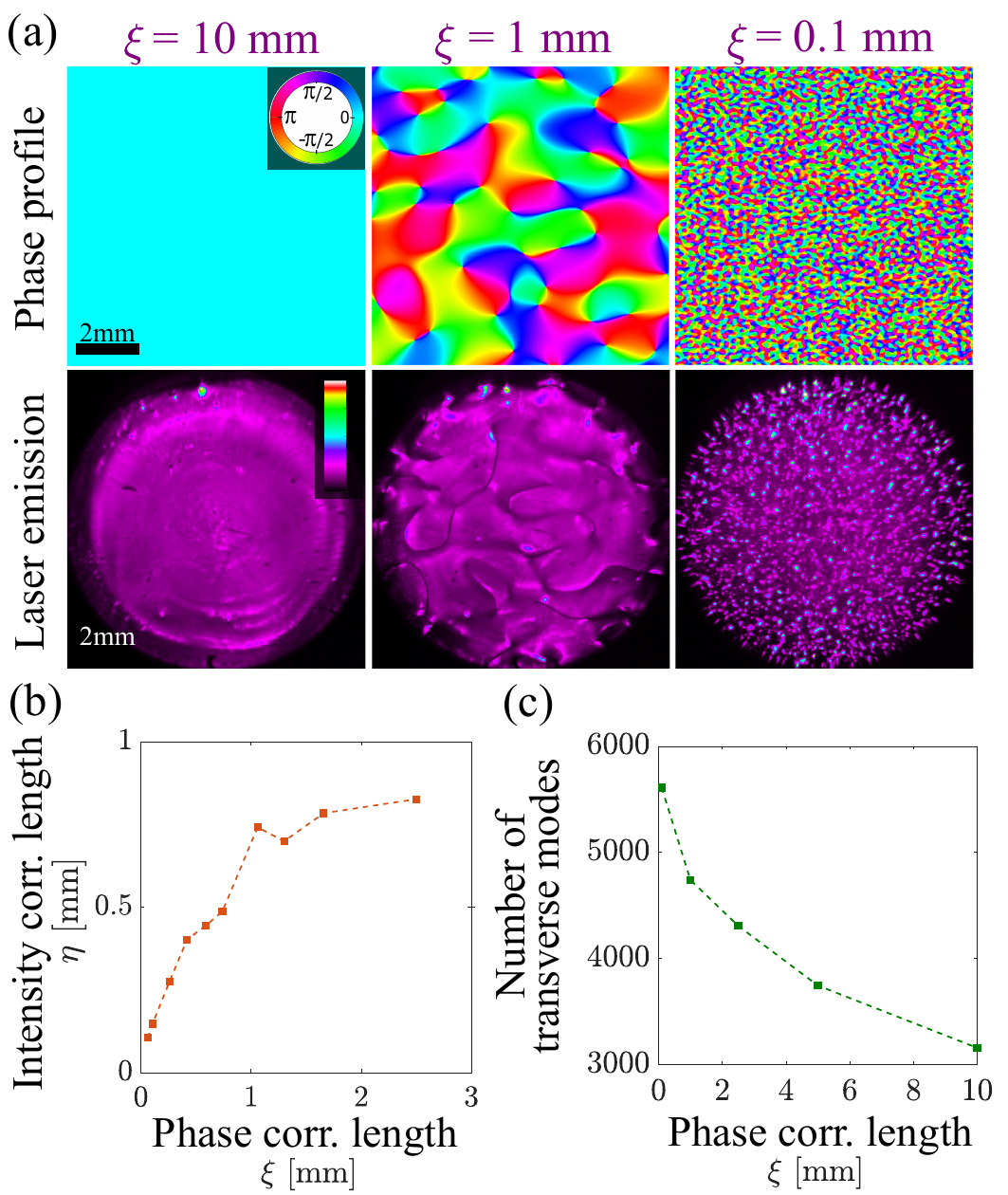}
	\caption{{Fragmented emission of DCL with random phase fluctuations.} {(a)} Random phase profiles displayed on the SLM (top row), and corresponding emission intensity patterns at the DCL output coupler (bottom row). The pump power is fixed at twice of the lasing threshold for flat-phase SLM. Left column: a flat phase over the cross-section of the cavity ($\xi$ = 10 mm) leads to homogeneous, flat-top emission pattern.  Middle column: a random phase profile with $\xi$ = 1 mm segments the lasing modes into multiple domains. Right column: a random phase profile with $\xi$ = 0.1 mm breaks the emission into many bright spots that are spatially localized. {(b)} Spatial correlation length of lasing intensity $\eta$ increases with SLM phase correlation length $\xi$. The feature size of emission pattern follows the phase fluctuation length, until it saturates when $\xi$ approaches the cavity transverse dimension. {(c}) Number of independent transverse lasing modes $N$ increases as $\xi$ decreases, indicating random phase fluctuation facilitates many-mode lasing.}
	\label{fig:experiment}
\end{figure}

We apply a series of random phase profiles to the SLM, with the correlation length $\xi$ varying from 10 mm to 0.1 mm. 
Figure~\ref{fig:experiment}(a) shows the emission intensity distribution at the OC plane for $\xi$ = 10, 1, 0.1 mm at the pump power of 2.4 times the lasing threshold for flat phase. At $\xi$ = 10 mm (equal to the transverse dimension of the cavity), the emission is homogeneous and has a flat top profile (left column). As $\xi$ decreases, the emission pattern is segmented into multiple domains (middle column). A further reduction of $\xi$ to 0.1 mm breaks the emission into many bright spots, each corresponding to a lasing mode (right column). The neighboring lasing modes are mutually incoherent, as they do not interfere with each other~\cite{Supplemental}. 

To characterize the feature size of emission pattern, we compute the spatial correlation function of the intensity distribution $I(x,y)$ at the OC plane, $C_I(\Delta x, \Delta y) = \langle I(x, y) \, I(x+ \Delta x, y + \Delta y) \rangle_{x, y}$, and its FWHM gives the correlation length $\eta$~\cite{Supplemental}. Figure~\ref{fig:experiment}(b) is a plot of $\eta$ versus $\xi$. At small $\xi$, $\eta$ increases almost linearly with $\xi$, and saturates when $\xi$ becomes comparable to the cavity transverse dimensions. As $\xi$ varies over two orders of magnitude, the total emission power changes by merely 30\%~\cite{Supplemental}. 

Next, we estimate the number of transverse lasing modes as a function of the phase correlation length $\xi$. To this end, we place a static glass diffuser outside the DCL and record the speckle pattern produced by the laser emission passing through the diffuser. The intensity contrast $C$ of a time-integrated speckle pattern gives the number of independent transverse lasing modes $N = 1/C^2$~\cite{goodman2007speckle, cao2019complex}. 

Fig.~\ref{fig:experiment}(c) shows how $N$ evolves with $\xi$ at a constant pump power. As the phase correlation length $\xi$ decreases, the number of independent transverse lasing modes increases. This indicates that introducing disorder to a degenerate cavity facilitates many-mode lasing. As the characteristic length scale of disorder decreases, the fragmentation of lasing modes reduces their spatial overlap and suppresses their competition for gain. The decrease of nonlinear modal interaction is dominant over the increase of diffraction loss with disorder, allowing more modes to lase simultaneously at the same pumping level.  

To understand the effects of random phase fluctuations on transverse modes, we conduct a numerical simulation of a DCL with varying degree of disorder. The laser configuration and dimensions are identical to the experimental realization, with the exception that the simulated cavity has a one-dimensional (1D) transverse cross-section to reduce computing time~\cite{Supplemental}. We first investigate how the transverse modes in a passive cavity are modified by a random phase fluctuation. Experimentally the DCL suffers from optical aberrations, misalignment and thermal lensing effect, thus a slight deviation from perfect degenerate condition is incorporated to the numerical simulation~\cite{Supplemental}. We calculate the transverse spatial profile and quality factor of cavity resonances. Then we study the lasing modes using the steady-state ab-initio lasing theory (SALT)~\cite{PhysRevA.74.043822}. Since introducing random phase fluctuations in the cavity's transverse direction has little impact on the mode quantization in the longitudinal direction, any change of longitudinal mode profile caused by a transverse phase modulation is neglected~\cite{Supplemental}. The nonlinear modal interaction via gain saturation is characterized by the cross-saturation coefficient, 
\begin{equation}
\chi_{m n} \cong \left| \int \psi^2_m(x) |\psi_n(x)|^2 \,dx \right| \, ,
\end{equation}
for $m$-th and $n$-th transverse modes, where $\psi_m(x)$ and $\psi_n(x)$ denote their transverse field profiles~\cite{ge2010steady}.     

\begin{figure}[hbtp]
\centering
    \includegraphics[clip, trim = 0.0cm 0.0cm 0.0cm 0, width=1\linewidth]{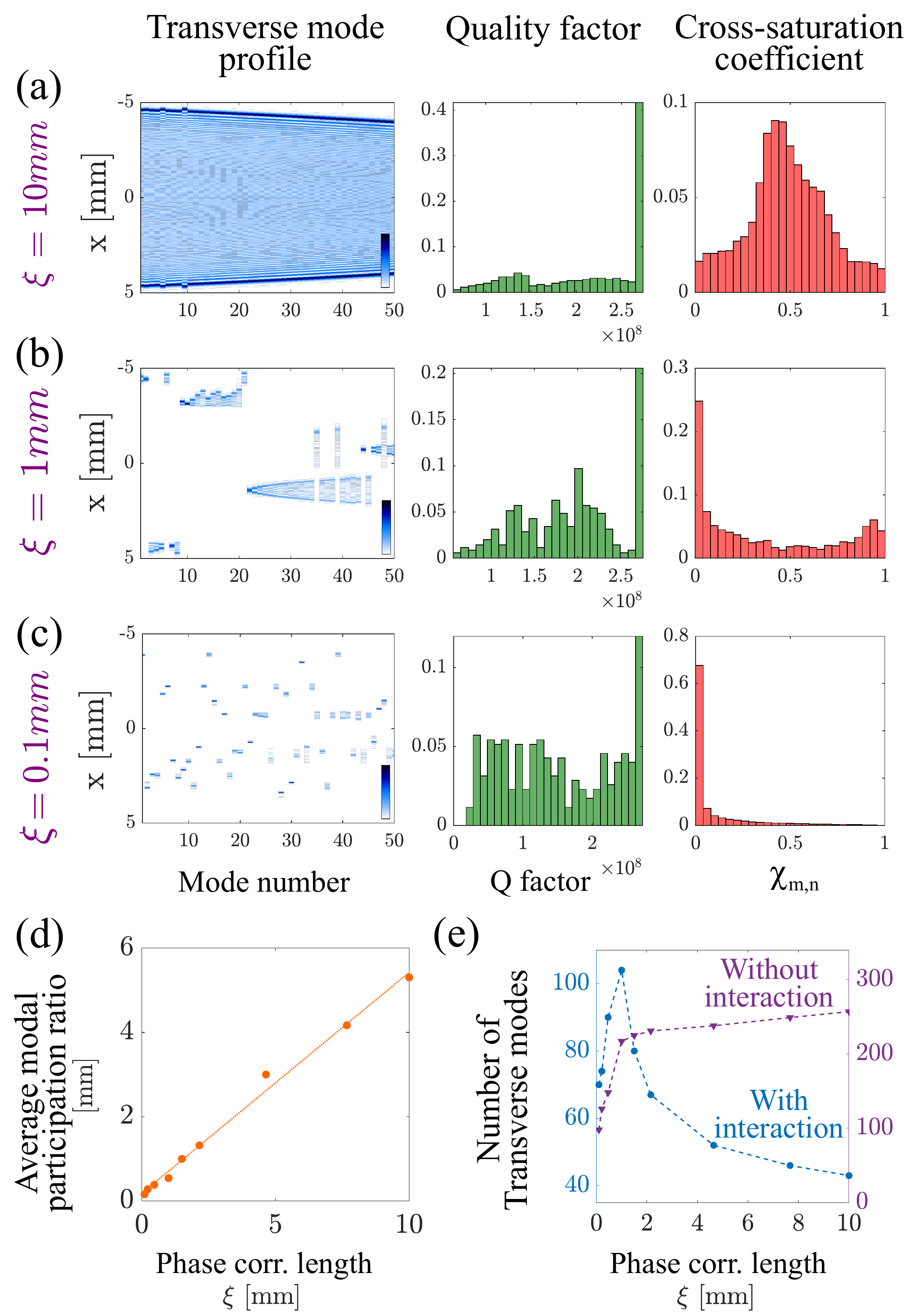}
\caption{{Suppression of modal interaction and $Q$ spoiling by disorder (simulation).} The left column in (a)-(c) shows the calculated 1D intensity profile of transverse modes in a slightly misaligned DCL. The center and right columns are distributions of quality factors and cross-saturation coefficients $\chi$. The random phase fluctuation length $\xi$ = 10 mm (a), 1 mm (b), and 0.1 mm (c).  
{(d)} Average mode size $\eta$ scales linearly with $\xi$. The solid line is a linear fit of slope = 0.52. 
{(e)} Number of transverse lasing modes as a function of $\xi$, with (blue circles) and without (purple triangles) gain saturation, at a constant pumping level of twice the lasing threshold with $\xi$ = 10 mm.}
\label{fig:simulation}
\end{figure}

With a flat phase on the SLM in Fig.~\ref{fig:simulation}(a), the transverse modes are spatially extended over the cavity cross-section. The distribution of their quality factors exhibits a narrow peak at the highest $Q$ value, indicating that the majority of transverse modes have similarly low lasing thresholds and tend to lase together. However, the large spatial overlap of these modes results in their strong competition for optical gain. The cross-saturation coefficients feature a wide distribution centered about $0.5$. We compute the number of lasing modes with gain saturation turned on and off. At the pumping level of $P = 2 P_0$, where $P_0$ is the threshold of the first lasing mode, the number of lasing modes decreases from 257 without modal interaction to 43 with modal interaction. This notable reduction reflects the important role played by nonlinear modal interaction.  

In Fig.~\ref{fig:simulation}(b), the SLM displays a random phase profile of correlation length $\xi$ = 1 mm, and the transverse modes shrink in size. They tend to cluster in regions with relatively smooth phase profile, avoiding the positions of abrupt phase change. The $Q$ distribution still features a narrow peak at the highest value, but the peak height is smaller, and more modes have lower $Q$ and higher lasing threshold. In contrast, the distribution of cross-saturation coefficients is peaked at the smallest value, and has a long tail extended to large $\chi$. The average cross-saturation coefficient is 5 times lower than that in Fig.~\ref{fig:simulation}(a), as a result of smaller spatial overlap between the transverse modes. At the pumping level of $2 P_0$, the number of lasing modes without interaction drops slightly to 217, while with interaction the number of lasing modes rises significantly to 104. This behavior indicates that the reduction of gain competition by the random phase fluctuation has a much stronger effect than the reduction of the $Q$ factors.

When the phase correlation length is reduced to $\xi$~=~0.1~mm in Fig.~\ref{fig:simulation}(c), the transverse modes become tightly confined with little overlap. This leads to a significant suppression of modal interaction, where the distribution of cross-saturation coefficients features a higher peak at the smallest value and a much shorter tail than that in Fig.~\ref{fig:simulation}(b). The $Q$ distribution is further extended to lower values, due to increased diffraction loss of highly localized modes. Consequently, both the number of lasing modes with and without interaction are reduced, the former to 70 and the latter to 98 at the same pumping level of $2 P_0$. 

Next we quantify the relation between the transverse mode dimension $\rho$ and the phase correlation length $\xi$. The size of $m$-th transverse mode is estimated from the participation ratio of its transverse intensity profile $|\psi_m(x)|^2$ as~\cite{kramer1993localization}:
\begin{equation}
\rho_m = \frac{[\int |\psi_m(x)|^2 \, dx]^2}{ \int |\psi_m(x)|^4 \, dx} \, .
\end{equation}
Figure~\ref{fig:simulation}(d) shows the average size of transverse modes $\bar{\rho} = \langle \rho_m \rangle_m$ as the phase correlation length $\xi$ varies over two orders of magnitude. The linear scaling of $\bar{\rho}$ with $\xi$ indicates that the transverse modes adapt to the random phase fluctuation and become localized accordingly in qualitative agreement with the results in Fig.~\ref{fig:experiment}.

Finally, we compare the number of transverse lasing modes with and without nonlinear interaction. If gain saturation is neglected (without interaction), the number of lasing modes depends only on their loss ($Q$ factor). As $\xi$ gradually decreases from 10 mm, the transverse modes start shrinking, and the diffraction loss becomes stronger. The reduction in $Q$ factors leads to higher lasing thresholds. As the pumping level is fixed to $2 P_0$, the number of lasing modes drops gradually. Once the transverse mode size is below the diffraction limit set by the numerical aperture of the cavity, a sharp increase of diffraction loss results in a sudden decrease in the number of lasing modes, as seen in Fig.~\ref{fig:simulation}(e). When gain saturation is included (with interaction), the trend is reversed: the number of lasing modes grows as $\xi$ is reduced from 10 mm to 1 mm. This is attributed to the reduced modal competition for gain, as the transverse modes are fragmented by random phase fluctuation. Once $\xi$ is shorter than $1 mm$, the dramatic increase of diffraction loss becomes dominant over the decrease of nonlinear modal interaction, and the number of lasing modes decreases accordingly [Fig.~\ref{fig:simulation}(e)]. However, the decrease in number of lasing modes with interaction is smaller than without interaction, indicating that the suppression of gain competition remains effective in allowing more transverse modes to lase.
Experimentally the drop of the number of lasing modes at very small $\xi$ is not observed, as a further decrease of $\xi$ below 0.1 mm would make the lasing modes so small that their intense emission might damage the SLM. A quantitative comparison between experimental data and numerical results is not possible, as the dimensions of the cavity cross-section differs and cavity imperfections cannot be accurately measured and adopted in the numerical simulation.


In conclusion, we demonstrate an efficient method of tuning nonlinear interaction of lasing modes over a wide range. By introducing random phase fluctuation into a degenerate cavity laser (DCL), the transverse modes are fragmented spatially to avoid the lossy regions of abrupt phase variation. The characteristic scale of phase fluctuation is varied over two orders of magnitude, and the transverse mode size follows.  The reduction of their spatial overlap suppresses modal competition for gain, resulting in an increase of the number of lasing modes, despite of $Q$ spoiling.   Contrary to typical laser cavities with fixed geometry, the spatial light modulator placed inside a DCL allows controlling the spatial structures and nonlinear interactions of thousands of lasing modes on-demand. Our approach is accurate, flexible and robust, providing a versatile platform for experimental study of many-body phenomena.

\begin{acknowledgments}
     The work done at Weizmann is supported by the Israel Science Foundation (ISF) under Grant No. 1881/17. The study performed at Yale is supported by the US Office of Naval Research (ONR) under Grant Nos. N00014-21-1-2026 and N00014-20-1-2197.
     The authors acknowledge the computational resources provided by the Yale center for research computing (Yale HPC).
\end{acknowledgments}


\bibliographystyle{apsrev4-2}
\bibliography{bibliography.bib} 

\end{document}